\documentclass[11pt]{article}

\usepackage{graphicx}
\usepackage{graphicx,psfrag}
\usepackage{amsmath,amssymb,amsthm}
 
\def \G {\Gamma}

\def \E {{\bf  \, E  }}

\def \Tr {{\hbox{\,Tr}}}

\def \bC {{\mathbb C}}

\def \bN {{\mathbb N}}

\def \bR {{\mathbb R}}

\def \CA {{\cal A}}

\def \CC {{\cal C}}

\def \CH {{\cal H}}

\def \CT {{\cal T}}

\def \CW {{\cal W}}

\def \CI {{\cal I}}

\def \CJ {{\cal J}}

\def \G{{\Gamma}}

\def \a {\alpha}

\def \b {\beta}

\def \l {\lambda}
\def \r {\rho}

\def \s {\sigma}

\def \d {\delta}

\def \Th {\Theta}

 \def  \e {\eta}

\def \rt {{\mathrm t}}

\def \det {\hbox{det}}

\def \tA {\tilde A}
\def \tB {\tilde B}

\begin{document}
\title{ On Eigenvalue Distribution of  Random Matrices of Ihara Zeta Function of 
Large Random Graphs\footnote{{\bf MSC:} 
05C50, 05C80, 15B52,  60F99
}
}

\author{O. Khorunzhiy\\ Universit\'e de Versailles - Saint-Quentin \\45, Avenue des Etats-Unis, 78035 Versailles, FRANCE\\
{\it e-mail:} oleksiy.khorunzhiy@uvsq.fr}
\maketitle
\begin{abstract}
We consider the ensemble of 
 real symmetric random matrices 
$H^{(n,\rho)}$ obtained from 
the  determinant form of the 
Ihara zeta function of random graphs that have $n$ vertices with the edge probability $\rho/n$.
We prove that the normalized  eigenvalue counting function of $H^{(n,\rho)}$ weakly converges 
in average as $n,\rho\to\infty$ and $\rho= o(n^\a)$ for any  $\a>0$ 
to a shift of the Wigner semi-circle distribution.  Our results support 
a conjecture that the large Erd\H os-R\'enyi  random graphs satisfy in average the weak graph theory Riemann Hypothesis.
\end{abstract}
\vskip 0.3cm



\section{Ihara zeta function of  graphs and random matrices}

The Ihara zeta function (IZF) 
associated to a finite connected graph $\G= (V,E)$ is defined at $u \in \bC$, for $\vert u\vert $ sufficiently small,    by 
$$
Z_\G(u) = \prod _{[C]}\  (1- u^{\nu(C)})^{-1},
\eqno (1.1)
$$
where the product runs over the equivalence classes of primitive closed backtrack-less, tail-less
cycles $C= (\a_1,\a_2, \dots, \a_{l}=\a_1)$ of positive length $l$ in $\G$, $\alpha_i\in V$
and \mbox{$\nu(C)=l-1$} is the number of edges in $C$ \cite{Te}.  
Being introduced by Y. Ihara \cite{I} in the algebraic context, 
IZF represents  now an intensively developing subject of combinatorial graph theory with applications in the number theory and the spectral theory
 (see e.g. \cite{HST,Te} and references therein); 
 it has also been studied in various other aspects, in particular 
 in relations with the heat kernels on graphs \cite{CJK}, 
 quantum walks on graphs \cite{RAE}, certain theoretical physics models 
 \cite{ZXH}.

Ihara's theorem \cite{I} proved first for  $(q+1)$-regular graphs  $\G$ says that the IZF (1.1)
is the reciprocal of a polynomial and that for sufficiently small $\vert u\vert$
$$
Z_\G(u)^{-1} = (1-u^2)^{r-1}\,  \det (I+u^2 (B-I) - uA),
\eqno (1.2)
$$
where $A = (a_{ij})_{i,j=1,\dots, n}$ is the adjacency matrix of $\G=\G_n$, $n$ 
is the number of vertices of $\G_n$, $B = diag( \sum_{j=1}^n a_{ij})_{i=1,\dots,n}$ and 
$r-1 = \Tr (B-2I)/2$. 
 This statement
 has been proven also for possibly irregular finite graphs (see \cite{Ba,ST} 
 for the combinatorial proofs of (1.2) and references related). The right-hand side of (1.2) represents an entire function; this means that  $Z_{\Gamma}(u)$ has a meromorphic continuation to the whole complex $u$-plane. Note also that $r-1$ can be expressed in terms of the Euler characteristic
 of the graph because $\Tr (B-2I)/2 = \vert E\vert - n$, where $\vert E\vert$ is the total number of \mbox{edges of $\G$.}
 
 While the Ihara's determinant formula (1.2) gives a powerful tool in the studies of  the Ihara zeta function,
 the explicit form of $Z_\G(u)$ can be computed  
 for relatively narrow families of  finite graphs. 
Regarding the case  of infinite graphs,  the definition of the Ihara zeta function  represents 
 an
 important problem that  requires 
 a number of additional  restrictions and assumptions
 (see, in particular, \cite{CMS,GIM-2,GZ,LPS}). 
 In the most cases, the graphs under consideration have a bounded vertex degree
 (in particular, regular or essentially regular graphs).

  A complementary approach  is represented by a  stochastic point of view,
 when the graphs are chosen at random from a set of all possible graphs on $n$ vertices.  
 This  description "in the whole"  naturally leads to the limiting transition  of infinitely increasing
 dimension of the graphs, $n\to\infty$.
 Certain aspects of 
 large random $d$-regular graphs have been studied in \cite{F}.
 The present note is related with 
  the Ihara zeta function (1.1) of random graphs 
  whose average vertex degree $\rho$  infinitely increases in the limit $n \to\infty$.

 \vskip 0.1cm
 Let us consider an ensemble of $n\times n$  real symmetric matrices $A^{(n,\r)}$ whose 
entries  are determined  by a collection of jointly independent Bernoulli random variables
$\CA_n^{(\r)} = \{a^{(n,\r)}_{ij}, 1\le i\le j\le n\}$ such that
$$
\left( A^{(n,\r)}\right)_{ij} = a^{(n,\r)}_{ij} = 
 \begin{cases}
  1-\delta_{ij} , & \text{with probability ${\displaystyle \r\over \displaystyle n}\, $} , \\
0, & \text {with probability $1-{\displaystyle \r\over \displaystyle n}\, $,}
\end{cases} 
\eqno (1.3)
$$
where $\d_{ij}$ is the Kronecker $\d$-symbol and $0<\r<  n$, $\rho \in \bN$. 
The adjacency matrices $\{A^{(n,\r)}\}$
 represent  the ensemble of random  graphs $\{\Gamma^{(n,\rho)}\}$ that can be  referred to as the
 Erd\H os-R\'enyi  random graphs  \cite{B}.
 With this definition in hands, one can determine  the random Ihara zeta function  $Z_{\G^{(n,\rho)}}$ 
 and to study it in the limit of infinite graph dimension, $n\to\infty$. 
 In the present note, we consider the 
 asymptotic regime of sparse random graphs, 
 when  $1\ll \rho \ll n^\a$ for any $\a>0$.  
\vskip 0.1cm 

Passing to  the  normalized logarithm of (1.2), we get  the following relation, 
$$
- {1\over n} \log Z_{\G^{(n,\r)}}(u)= {1\over 2n}\, \Tr\left( B^{(n,\r)}-2I\right)\, \log (1-u^2)  
$$
$$+ 
{1\over n}
\log \det \left((1-u^2)I +u^2 B^{(n,\r)}-uA^{(n,\r)}\right),
\eqno (1.4)
$$
where
$$
\left( B^{(n,\r)}\right)_{ij} = \d_{ij} \sum_{l=1}^n a^{(n,\r)}_{il}.
$$
Regarding the first term of the right-hand side of (1.4)
$$
\Th^{(n,\r)}(u) = {1\over 2n}\Tr\left( B^{(n,\r)}-2I\right)\, \log (1-u^2),
$$
it is easy to compute its mathematical expectation 
 $\E \Th^{(n,\r)}(u)$ with respect to the measure generated by the family $\CA_n^{(\rho)}$ (1.3),
 $$
 \E \Th^{(n,\r)}(u) = \left( {n-1\over 2n} \r-1\right) \ln (1-u^2) \to \left( {\r\over 2} -1\right) \ln (1-u^2), \quad n\to\infty.
 \eqno (1.5)
$$
The last expression shows that  to obtain a finite value of 
$\E \Th^{(n,\r)}(u)$ in the 
limit \mbox{$n,\r\to\infty$}, one has to rescale the parameter $u=u_\r$  as follows,
$$
u_\r^2 = {v^2\over  \r} .
\eqno (1.6)
$$
Then the last term  of (1.4) takes the form
$$
\Xi^{(n,\r)}(v) = {1\over n}
\log \det \left((1-v^2/\r)I +H^{(n,\r)}(v)\right), 
\eqno (1.7)
$$
 where 
 $$
 H^{(n,\r)}(v) =  {v^2\over \r} B^{(n,\r)} - {v\over \sqrt \r} A^{(n,\r)}.
 \eqno (1.8)
 $$
   \vskip 0.1cm 
   
   The presence of the factor  $ \r^{-1/2}$ in front of  $A^{(n,\r)}$ 
is fairly natural from the point of view of the spectral theory of large random
matrices. The normalization of $B^{(n,\rho)}$  by  $v^2/\rho$ 
is less common and  arises because of the condition  (1.5). 
Therefore it would be  natural to say that 
 $\{H^{(n,\rho)}(v)\}$ is the random matrix ensemble  of the Ihara zeta function of random graphs.
It is interesting to note that similar rescaling of the spectral parameter $u^2$  (1.6)
by the vertex degree $q+1$ is  needed when instead of  $\{ \Gamma^{(n,\rho)}\}$
the ensemble of $(q+1)$-regular random graphs is considered  in the limit of infinite $q$ \cite{McKay}. 

\vskip 0.1cm 
Let us  note that in the case of  $u=1$, the last term of the right-hand side of (1.4)
gives   the discrete version of the Laplace operator $\Delta^{(n,\rho)}_\Gamma = B^{(n,\rho)} - A^{(n,\rho)}$ determined on 
the  graph $\Gamma^{(n,\rho)}$. 
The spectral properties of $\Delta^{(n,\rho)}_\Gamma$ of large random graphs 
have been studied in a number of works (see e. g.  \cite{C-O,KKM}). 
The  eigenvalue distribution  of a version of the graph's Laplacian 
given by $\rho^{-1/2} \Delta_{\Gamma}^{(n,\rho)}$ for finite and infinite values of $0<\rho<n$ 
was  studied in the limit $n\to\infty$ in \cite{KSV}. 
Let us stress  that this matrix $\rho^{-1/2} \Delta_{\Gamma}^{(n,\rho)}$ 
is again different from (1.8).
The "zeta-function" normalization induced   by (1.5) 
essentially changes the properties of the ensemble with 
respect to the graph's Laplacian.
In particular, 
in contrast to $\Delta_{\Gamma^{(n,\rho)}}$ the matrix  
$H^{(n,\rho)}(v)$ with given $v>0$ (1.8) is no more  positively determined
for large values of  $\rho$, even the finite ones. 

Summing up, we can say that the random matrix ensemble $H^{(n,\rho)}(v)$ (1.8) is the new one
that  
up to our knowledge has not been studied.
In the present note we consider  (1.8) with   
 $v\in \bR  $ and 
study the  eigenvalue distribution of real symmetric random matrices $H^{(n,\rho)}(v)$ 
in the limit 
$$
n,\rho\to \infty, \quad \rho = o(n^\a), \ {\hbox{for any }}  \a>0
\eqno (1.9)
$$ 
that we denote by $(n,\rho)_\a\to\infty$. 
 Our main proposition  is that the normalized eigenvalue counting function of $H^{(n,\rho)}(v)$
 weakly converges in average to the well-known Wigner semicircle distribution shifted by $v^2$. 
 This statement is proved in Section 2. In Section 3, we discuss
 our results in relation with  the limiting values 
 of the correspondingly re-normalized  Ihara zeta function for complex $v\in \bC$.

  \section{Limiting eigenvalue distribution of $H^{(n,\rho)}(v)$}
  Let us rewrite  definition (1.8) in the form 
  $$
  H^{(n,\r)}(v)=  v^2 \tB^{(n,\r)} - v \tA^{(n,\r)}, \quad v\in {\bR}.
  \eqno (2.1)
  $$
 Denoting the eigenvalues of $H^{(n,\r)}(v)$  by $\l_1^{(n,\r)}(v)\le \dots\le  \l_n^{(n,\r)}(v)$, 
one introduces the normalized eigenvalue counting function,
$$
\s_v^{(n,\r)}(\l) = {1\over n}  \#\left\{ j: \, \l_j^{(n,\r)}(v) \le \l\right\}, \quad \lambda \in \bR.
\eqno (2.2)
$$
The moments of this measure satisfy the following relation, 
$$
M_{k}^{(n,\r )} (v) 
= \E \left( {1\over n}\Tr \left(H^{(n,\r)}(v)\right)^k\right) = \int_{-\infty}^{+\infty} \, \l^k d\bar \s_v^{(n,\r)}(\l), 
\quad k= 0, 1, 2, \dots, 
\eqno (2.3)
$$
where  $\bar \s_v^{(n,\r)}$ represents the averaged eigenvalue counting function,  
$\bar  \s_v^{(n,\r)} = \E \s_v^{(n,\r)}    $.
The  main  result of the present note  is as follows.

\vskip 0.3cm

{\bf Theorem 1.} {\it For any given $k\in \bN$, the averaged moment (2.3) converges in the limit (1.9),
$$
\lim_{(n,\r)_\a\to\infty} M_k^{(n,\r)}(v) = \tilde m_k(v) = 
 \begin{cases}
 v^{4l+2} \  {\displaystyle \sum_{p=0}^l}  \ {\displaystyle 1\over \displaystyle v^{2p}} \,
{\displaystyle {2l+1 \choose 2p}}\,  \rt_p, & \text{if $k=2l+1$}, \\
v^{4l} \  {\displaystyle \sum_{p=0}^l}  \ {\displaystyle 1\over \displaystyle v^{2p}} \,
{\displaystyle {2l \choose 2p}}\,  \rt_p, & \text {if $k=2l$,}
\end{cases}
\eqno (2.4) 
$$
where $l= 0, 1, 2, \dots$ and 
$$
\rt_p = {\displaystyle {(2p)!\over p!\, (p+1)!}},\quad  p=0,1,2,\dots
\eqno (2.5)
$$
are the Catalan numbers.  }

\vskip 0.3cm
{\it Proof.} 
To study the moment $M_k^{(n,\rho)}$, we consider 
the product
$$
L_k^{(n,\rho)}(\bar P, \bar Q) = {1\over n} \sum_{i_0=1}^n 
\ \E \left( \tA^{p_1} \, \tB^{q_1} \cdots \tA ^{p_s} \tB ^{q_s} \right)_{i_0i_0} , 
\eqno (2.6)
$$ 
  where $\sum_{i=1}^s p_i = P$, \mbox{$\sum_{i=1}^s q_i = Q$},  $P+Q=k$,
  and where we denoted  $\bar P = (p_1, \dots, p_s)$ and $\bar Q = (q_1, \dots, q_s)$. 
In (2.6), we assume \mbox{$p_2\ge 1, \dots, p_s\ge 1 $}, \mbox{$q_1\ge 1, \dots, q_{s-1} \ge 1$} and 
$p_1\ge 0, q_s\ge 0$.  In what follows, we  omit the bars over $P$ and $Q$ when no confusion can arise.

 We study the limiting behavior of variables $L_k^{(n,\rho)}(\bar P, \bar Q)$ 
 with the help of the diagram technique  close to that
developed in  \cite{KSV}.  
We consider the product 
  $$
 \left( \tA^{p_1}  \cdots  \tB ^{q_s} \right)_{i_0i_0}  = \sum_{i_0, i_1, \dots i_{p_s}=1}^n \big(\tA^{p_1}\big)_{i_0 i_{p_1}} 
 \big( \tB^{(q_1)}\big)_{i_{p_1} i_{p_1}}\cdots \big(\tA^{p_s}\big)_{i_{p_1+\dots+p_{s-1}}+1, i_{0}}
  \big( \tB^{(q_s)}\big)_{i_{0} i_{0}}
$$
 as a sum over  the  generalized 
trajectories $\big(\CI_{p_1}^{(1)}, \CJ_{q_1}^{(1)}, \dots, \CI_{p_s}^{(s)}, \CJ_{q_s}^{(s)}\big)$,
 where the closed trajectory of $P$ steps 
is given by 
$$
\CI_P= (\CI_{p_1}^{(1)}, \CI_{p_2}^{(2)}, \dots , \CI_{p_s}^{(s)}) = (i_0, i_1, \dots, i_{p_1}, i_{p_1+1}, \dots, i_{p_1+p_2}, \dots, i_{P-1}, i_0)
\eqno (2.7)
$$
and $\CJ^{(k)}_{q_k} = (j_1^{(k)}, \dots, j_{q_k}^{(k)})$, $k=1, \dots, s$. 

Regarding (2.7), we associate to $i_0$ a root vertex 
$\alpha$ and draw the new vertex $\beta, \gamma, \delta, \dots$ each time when we see a value of $i_l$ 
that is not equal to one of the values previously seen. 
We draw the blue edges that correspond to the steps of $\CI_P$ and 
get a graphical representation of the trajectory $\CI_P$ by 
 a closed chain $G_P$ with $P$ blue edges. 
Clearly,    $G_P$ is a connected multi-graph with the root vertex $\alpha$. 
We denote by $\beta_1, \dots, \beta_h$ the vertices of $G_P$ that correspond 
to different values of variables $i_0, i_{p_1}, i_{p_1+p_2}, \dots, i_{p_1+p_2+ \dots+{p_{s-1}}}$, $1\le h\le P$.

The $q_1$-plet $\CJ_{q_1}^{(1)}$ 
can be represented by a set of $q_1$ oriented red edges  $(\b_1, \gamma_l)$ with $1\le l\le q_1$.
We can put a flesh at the head vertex $\gamma_l$. 
If the value of $j^{(1)}_r$ does not coincide with any element of $\CI_P$ (2.7), then
we say that the corresponding vertex $\gamma_r$ is the red one.  In the opposite case we have $\gamma_r \in V(G_P)$. 
Then we construct,  by the same procedure, a representation of the remaining parts 
$\CJ^{(2)}_{q_2}, \dots, \CJ^{(s)}_{q_s}$ and get a multi-graph
that we denote by 
$\CH(\bar P, \bar Q) = G_{ \bar P}\uplus F_{\bar  Q}$. We say that $G_{\bar P}$ and $F_{\bar  Q}$ represent the blue and the red sub-graphs
of the diagram $\CH(\bar P, \bar Q)$, respectively.

Let denote by $\bar G = \bar G_{\bar P}$ and 
$\bar F = \bar F_{\bar Q}$  the simple graphs associated 
with $G= G_{\bar P}$ and $F = F_{\bar Q}$, respectively
and consider
a part  $\bar F'_{\bar Q}$ of the red sub-graph 
of $\bar \CH(\bar P, \bar Q)$ 
such that there is no edge of $E(\bar  F'_{\bar Q})$ 
that  coincide with the elements of $E(\bar G_{\bar P})$.  
We denote by $V_r(\bar F')$ the set of red vertices 
with fleshes of   $\bar F'_{\bar Q}$ and write that  
$\bar F''_{\bar Q} = \bar F_{\bar Q} \setminus \bar F'_{\bar Q}$. 
It is clear that
$$
\Pi_{A,B}({\cal I}_P,{\cal J}_Q) = 
\left( {\rho\over n}\right)^{\vert E(\bar G)\vert + \vert E(\bar F')\vert}
$$
where $\bar F' = \bar F'_{\bar Q}$.  Also it is easy to see that
$$
\vert \CC(G_{\bar P}\uplus F_{\bar Q})\vert =
n(n-1)\cdots (n- \vert V(\bar G)\vert  - \vert V_r(\bar F')\vert +1 ) = n^{ \vert V(\bar G)\vert  + \vert V_r(\bar F')\vert} (1+o(1)), \ n\to\infty.
$$
Then
$$
 {1\over n \rho^{P/2+Q}} \  
\sum_{\{\CI_P, \CJ_Q\}\in \CC(G_{\bar P} \uplus F_{\bar Q})} 
\ \Pi_{A,B}({\cal I}_P,{\cal J}_Q)
$$
$$= {\rho^{\vert E(\bar G)\vert-P/2}\over n^{\vert E(\bar G)\vert - \vert V(\bar G)\vert +1} }
\cdot  {\rho^{\vert E(\bar F')\vert - \vert E(F')\vert - \vert E(F'')\vert}
 \over n^{\vert E(\bar F')\vert - \vert V_r(\bar F')\vert }  }(1+o(1)), 
 \ n\to\infty,
\eqno (2.8) 
$$
where  $F'' = F''_{\bar Q}$. 

By using  (2.8), it is not hard to prove   that the terms of (2.6) that do not vanish in the 
limit 
$(n, \rho)_\a\to\infty$ (1.9) 
have the diagrams that satisfy inequality 
$$
\vert E(\bar G)\vert - \vert V(\bar G)\vert   +\vert E(\bar F')\vert - \vert V_r(\bar F')\vert +1 \le 0.
\eqno (2.9)
$$ 
Indeed, if (2.9) is not satisfied, then the right-hand side of (2.8) gets a factor $n^{-k}$, $k\ge 1$ that
suppress any power of $\rho$. The following two conditions are also necessary to have a non-zero limit:
$ \vert E(\bar G)\vert \ge  P/2$ 
and 
 $$
 \vert E(\bar F')\vert - \vert E(F')\vert
 -\vert E(F'')\vert \ge 0.
\eqno (2.10)
$$ 
Let us denote a diagram that verifies 
these three conditions  by 
$\tilde \CH(\bar P,\bar Q) = \tilde G_{\bar P}\uplus \tilde F_{\bar Q}$.
Due to the Euler relation for the planar embedding of connected graphs, the only equality 
is possible in (2.9) and this  means that the simple graph 
$\tilde  G\uplus \tilde F'$ is given by a tree. 
This tree is a plane rooted tree such that its blue sub-graph $\tilde G$  is also a tree with $P/2=p$ edges.  
Indeed, let us  denote  by $(\delta_1, \epsilon_1)$ the first leaf of the tree $\tilde G$. Then in the trajectory $\CI_P$ (2.7)
 there is a couple of steps $(i_{l-1}, i_l),(i_l, i_{l+1})$ such that $i_{l-1}=i_{l+1}$. 
 Removing this couple, we get the reduced 
 trajectory $\CI'_{p-2}$ such that  its representation $\tilde G$ is again a tree. Proceeding by recurrence, 
 we see that $P$ is pair, $P= 2p$ and that $\vert E(\tilde G)\vert \le P/2$. Thus,  $p= \vert E(\tilde G)\vert$.

Let us note that given a graph $\tilde G_{\bar P}$ that is
 a rooted tree $\CT_p$, one can easily restore  the original  multi-graph $\G_{\bar P}$ 
 and corresponding sequence of vertices $\CW_{2p}$
by considering the chronological run over the tree $\CT_p$.
In this case the walk $\CW_{2p}$ is such that $\G_{\bar P}$ is a multi-graph,
where each couple of vertices $\{\gamma,\delta\}$ is joined by either zero edges or exactly two edges - in there and back  directions, 
$(\gamma,\delta)$ and $(\delta,\gamma)$.

Regarding the red sub-graph $ F =  F'\sqcup F''$ of $\tilde \CH(\bar P, \bar Q)$, one can easily 
see that  $\vert E(\bar F')\vert - \vert E(F')\vert\in \{ 0, -1,- 2, \dots\}$ and that 
$\vert E(F'')\vert \in\{ 0, 1, 2, \dots \}$.  Then (2.10) is possible only when $\tilde F''$ is empty and 
when $\bar F' = \tilde F'$. The last relation means  
that   the red part $\tilde F' = \tilde F$ has no multiple edges.

It is known  that the number of rooted trees
of $p$ edges is given by the Catalan number $\rt_p$. The positions and numbers of red edges being 
determined by $(\bar P, \bar Q)$, we deduce from (2.8) that  the following relation is true,
$$
L_k^{(n,\rho)}(\bar P, \bar Q) = \rt_p v^{2Q+p}  (1+ o(1)), \quad (n,\rho)_\a\to\infty.
\eqno (2.11)  
$$
Returning to the moments (2.3) and regarding the trace
$$
{1\over n} {\bf E} \left( \Tr H_n^k \right) = {1\over n} \sum_{i_0=1}^n {\bf E}  \big(\underbrace{{H_n\, H_n\cdots H_n}}_{k\ times}\big)_{i_0i_0},
$$
we have to choose $2p$ elements $H_n$ of the last product that will be represented by $-\tA$.
This can be done in ${\displaystyle {k\choose 2p}}$ ways and this choice determines
uniquely the $s$-plets $\bar P$, $\bar Q$.  
Combining this observation with 
(2.11), we  get relation (2.4).
Theorem 1 \mbox{is proved.  $\Box$}

\vskip 0.3cm

It is known that the Catalan numbers $\rt_k$ verify the recurrence
 $$
 t_{k+1} = \sum_{j=0}^k \, \rt_{k-j} \rt_j, \ k\ge 0 \quad {\hbox{and}} \quad  \rt_0=1,
 \eqno (2.12)
 $$
 and that the family of moments 
 $$
 v^{2p} \rt_p = \int_{\bR} \l^{2p} d\mu_v(\l), \quad p= 0, 1,2,\dots
 $$ 
 uniquely determines an even
measure $\mu_v$ with  the  density
$$
{d\mu_v \over d\l} = \mu_v'(\l) ={1\over 2\pi v^2}
\begin{cases}
 \sqrt{4v^2 - \l^2}
  , & \text{if  $\lambda \in [-2v,2v] $}, \\
0, & \text {otherwise,}
\end{cases} 
  \eqno (2.13)
$$
known in the spectral theory of random matrices as the semi-circle or the Wigner distribution \cite{W}.
It follows from  (2.4) that the limiting moments $\tilde m_k(v)$ can be represented as follows,
$$
\tilde m_k(v) = \int_{\bR} (v^2 + \l)^k\, d\mu_v(\l), \quad k=0,1,2,\dots
\eqno (2.14)
$$
and therefore the corresponding measure $\tilde \s_v$ 
such that $\tilde m_k(v) = \int \lambda ^k d\tilde \s_v(\lambda)$ is given by a shift of the semi-circle distribution (2.13),
$$
\tilde \s'_v(\l) =  {1\over 2\pi v^2} \sqrt{4v^2 - (\l-v^2)^2}, \quad \vert \l-v^2\vert\le 2\vert v \vert.
\eqno (2.15)
$$
It follows from (2.15) that 
$$
\tilde m_k(v) \le (2v +v^2)^k, \quad k\ge 0.
$$
The family of moments $\{ \tilde m_k(v)\}$ satisfies the Carleman condition
and therefore the measure $\tilde \s_v$  is uniquely determined. 
Thus, Theorem 1 implies the weak convergence in average 
 of measures $\s_v^{(n,\rho)}$    to  $\tilde \s_v$;
this means that  for any continuous bounded function $f: \bR\mapsto \bR$ the following is true,
$$
\lim_{n,\r\to\infty} \int_{\bR} f(\l ) \, d\bar \s_{v}^{(n,\r)}(\l) = 
\int_{\bR} f(\l ) \, d\tilde \s_v(\l).
\eqno (2.16)
$$
In this connection it should be noted that the generating function 
$$
f(\xi) = - \sum_{k=0}^\infty {1\over \xi^{k+1}} v^{2k} t_k = \int_{-2v}^{2v} {d\mu_v(\lambda)\over \lambda - \xi}
$$
verifies the following equation that can be easily deduced from (2.12),
$$
f(\xi ) = {1\over - \xi - v^2 f(\xi)}.
\eqno (2.17)
$$
Regarding the generation function $ g(\xi)$ of $ \tilde m_k(v)$ and using (2.14), we get equality
$$
g(\xi) = - \sum_{k=0}^\infty  {1\over \xi^{k+1}} \tilde m_k (v)= \int_{-\infty}^\infty {d \s_v(\lambda) \over \lambda - \xi}
= \int_{-\infty}^\infty {d\mu_v(\lambda - v^2)\over \lambda - \xi} = f(\xi-v^2).
$$
This means  that  $g(\xi)$ verifies the deformed version of (2.17),
$$
g(\xi ) = {1\over v^2 - \xi - v^2 g(\xi)}.
\eqno (2.18)
$$
Relation  (2.18)
shows that the numbers 
$\tilde m_k(v), k\ge 0$
 are determined by the  recurrence (cf. (2.12))
 $$
\tilde  m_{k+1} (v)= v^2\tilde  m_k(v) + v^2 \sum_{j=0}^{k-1} \, \tilde m_{k-1-j}(v) \, \tilde m_j(v), \quad k\ge 1
 \eqno (2.19)
 $$
 with the initial conditions $\tilde m_0(v) = 1$ and $\tilde m_1(v) = v^2$. 
 In the particular case of  $v^2 =1$
 the  first ten values of $\tilde m_k(1), k\ge 0 $ (2.19)  are given by 
 1, 1, 2, 4, 9, 21, 51, 127, 323, 835. One can compute this also with the help of (2.4).
 These numbers are  the Motzkin numbers \cite{OE}.
 

 \vskip 0.5cm
Using the diagram technique developed above, one can get the following  improvement of Theorem 1. 
 
 \vskip 0.1cm
 {\bf Theorem 2.} {\it Given $k$, the following relation holds,}
 $$
\lim_{(n,\rho)_\a\to\infty} \,  \rho  \left( M_k^{(n,\rho)} - 
\tilde m_k(v)\right) = R_k^{(1)}(v),
\eqno (2.20)
$$
 {\it where
 $$
 R_k^{(1)}(v)= v^{2k - 2p} \, \rt_p
$$
$$
\times \left(
 \sum_{p=0}^{\lfloor k/2\rfloor}   { k \choose 2p} {p(p-1)\over p+2} + 
 4 \sum_{p=0}^{\lfloor (k-1)/2\rfloor}    {k \choose 2p+1} p
 + \sum_{p=0}^{\lfloor (k-2)/2\rfloor}   {k\choose 2p+2} {4p+2\over p+2 }\right).
 \eqno (2.21)
 $$
 }

{\it Proof.}
Let us consider a diagram $\hat \CH(\bar P, \bar Q)$ such that corresponding sum over $\CC(\hat \CH(\bar P, \bar Q))$ represents
the term of the order $O(\rho^{-1})$ of $M_k^{(n,\rho)}$ in the limit  $n,\rho\to\infty$, $\rho=o(n)$. 
It follows from  (2.8) that the diagram $\hat \CH(\bar P, \bar Q)$ 
verifies condition (2.9) and therefore its blue part $\hat G$ 
together with its red part $\hat F'$ represents a plane rooted tree 
$\hat G\uplus \hat F' = \CT_r$. In these relations, we denoted by  $\hat G$ and  $\hat F$  simple graphs obtained from corresponding 
multi-graphs $G$ and $F$. The red part $\hat F'$ contains such red edges that do not coincide with edges of $\hat G$. 

The next consequence of (2.8)  is that the following condition is verified by $\hat \CH$,
$$
\vert E(\hat G)\vert -P/2 + \vert E(\hat F')\vert - \vert E(F')\vert 
-\vert E( F'') \vert = -1,
\eqno (2.22)  
$$
where $ F'' = F \setminus  F'$.
It is clear that  (2.22) is possible in one of the three
following situations: 
\vskip 0.2cm 
- either $\vert E(\hat G)\vert -P/2= -1$, $\vert E(\hat F')\vert - \vert E(F')\vert =0$ and $\vert E( F'') \vert=0$,

- or $\vert E(\hat G)\vert -P/2= 0$, $\vert E(\hat F')\vert - \vert E(F')\vert =0$ and $\vert E( F'') \vert=-1$,

- or $\vert E(\hat G)\vert -P/2= 0$, $\vert E(\hat F')\vert - \vert E(F')\vert =-1$ and $\vert E( F'') \vert=0$.

 \vskip 0.2cm 
In the first case  the chain of vertices of $\hat G_P$ is such that the corresponding walk $\hat \CW_{2p}$ 
generates a multi-graph 
 of the tree-type such that there exists one edge passed four times and that remaining $p-2$ edges are
 passed two times in there and back directions. The number of such walks 
 is given by the formula
 $$
\# \left\{ \hat \CW_{2p}\right\}  = {(2p)!\over (p-2)! \, (p+2)!} = \rt_p \, {p(p-1)\over (p+2)}
 \eqno (2.23)
 $$
 (see \cite{K2} for the proof). 
 Then the remaining $k-2p$ red edges are to be distributed over $2p+1$ instants of time.
This gives the factor ${k\choose 2p}$ 
in the right-hand side of (2.21).

\vskip 0.2cm
The second case describes a graph such that 
 the chain of vertices such that the blue diagram $\hat G_P$ is a rooted tree of $p= P/2$ edges passed two times
and there exists one  red edge of the form $(\beta,\gamma)$ such that $\gamma \in V(\hat G_P)$. 
It is not hard to see that obtain the corresponding chain, we have to consider  a tree
$\CT_p$, to choose an edge $e$ from it, to join one red edge $\e'$ to one or another side of $e$,
to choose the orientation of the flesh of $e'$ and to
distribute $k-2p-1$ red edges over $2p+2$ instants of time. This gives the second term 
of  the right-hand side of (2.21). 

\vskip 0.2cm
Finally, the third case determines  such   diagrams $\hat \CH$ that 
$\hat G_P$ is a tree of $p$ edges and there exist two red edges that make a multi-edge and the remaining 
 red edges are simple. Let us describe how to construct corresponding diagrams and chains of vertices (walks).
 We consider two red edges  of the form $e'= (\beta, \gamma)$ and $e'' =(\beta,\gamma)$.  Then we attach to the vertex 
 $\beta$ a tree $\CT_a$ that have $a$ edges and point out an instant of time that will represent the starting and ending point
 of the corresponding walk. This can be done in $2a+1$ ways. Finally, we attach to $\beta$ 
 a tree $\CT'_b$ of $b$ edges, $b=p-a$  in the way that it will represent the part of the walk  performed between the fist 
 and the second passages of $(\beta, \gamma)$ given by $e'$ and $e''$, respectively. 
 Now it remains to  distribute  $r-2-2p$ red edges among $2p+1+2$ 
instant of times. This gives the factor $ {k\choose 2p+2}$. Taking into account
elementary equality
$$
\sum_{a, b \ge 0 , a+b=p} (2a+1) \rt_a \rt_b = (p+1) \rt_{p+1},
\eqno (2.24)
$$
 we get the third term of the right-hand side of (2.21). 
 Relation (2.24) can be proved with the help of the generating function
 $f(\xi)$ \cite{K2}. Also one can observe that the right-hand side 
 of (2.24) represents the number of Catalan trees of  $p+1$ edges with 
 one marked edge; this gives the right-hand side  of (2.24).
  
 Theorem 2 is proved. $\Box$
 



\section{Applications to IZF}

Variable  (1.7) can be rewritten with the help of (2.2)
in the following form,  
$$
\Xi^{(n,\rho)}(v) = {1\over n} \log \det \left( 1_\rho + v^2 \tilde B^{(n,\rho)} - v \tilde A^{(n,\rho)}\right) = 
 \int_{-\infty}^{\infty} \log 
\big(
1_\rho +\lambda\big) d\s^{(n,\rho)}_v(\lambda),
\eqno (3.1)
$$
where we denoted $1_\rho = 1-v^2/\rho$.
Then the convergence of IZF for a sequence of graphs 
can be reduced to the question of the convergence
of the corresponding spectral measures.  This approach has been 
used for the first time in the studies of IZF of infinite regular graphs  in paper \cite{GZ}.

It is known that the normalized adjacency matrix $\tilde A^{(n,\rho)} = \rho^{-1/2} A^{(n,\rho)}$
of the Erd\H os-R\'enyi random graphs has all eigenvalues, excepting the maximal one, 
concentrated with probability 1 
in the limit $n,\rho\to\infty$ on the interval $[-2,2]$, while this maximal eigenvalue is of the order $\sqrt \rho$ \cite{FO,FK}.
Regarding (3.1) for the negative values of $v$, 
one can therefore expect that the limit
$$
\lim_{(n,\rho)_\a\to\infty} \Xi^{(n,\rho)} (v) 
= \Xi(v)
\eqno (3.2)
$$
exists with probability 1  for all   $v$ such that  \mbox{$-1/2< v<0$}. 
Moreover, observing  that $\tilde B^{(n,\rho)}$ is asymptotically  close to the unit matrix in the limit (1.9),
it would be natural to assume that convergence in average 
$$
\lim_{(n, \rho)_\a\to\infty} {1\over n} \E\big( \log Z_{\Gamma^{(n,\rho)}}(v/\sqrt \rho) \big) 
= {v^2\over 2}  -  \Xi(v),
\eqno (3.3)
$$
where 
$$
\Xi(v) = 
{1\over 2\pi v^2} \int_{-2\vert v\vert }^{2\vert v\vert } \log (1+\lambda ) \, \sqrt{ 4v^2 - (\lambda-v^2)^2}\, 
d\lambda
\eqno (3.4)
$$
holds for all  $v\in (-1,1)$. 
While (3.3) is close to the weak convergence result  \mbox{$\bar \s_v^{(n,\rho)} \to \tilde \s_v$} (2.16)
established by Theorem 1 above, 
 one   cannot  use it directly  because the function $\log (1+\lambda)$ is not bounded in the vicinity of $-1$.
 To justify (3.3), the detailed analysis of fine properties of eigenvalues of $\tilde H^{(n,\rho)}$ is needed.
This goes beyond  the scope of the present note.   

\vskip 0.2cm 
Expression  (3.4) can be rewritten in the form
$$
\Xi(v)=  {2\over \pi} \int_{-1}^1 \log\big(1+v^2 + 2v \nu\big) 
\sqrt{1 - \nu^2}\, d\nu, 
\eqno (3.5)
$$
where the last integral converges for any real $v$. 
Moreover, the right-hand side of (3.5) 
can be continued to a function  holomorphic in any   domain 
$$
C_\epsilon = \{ v: \ v\in \bC, \vert v\vert \le 1-\epsilon\}, \quad 0< \epsilon <1.
$$
Thus, assuming that (3.2) and (3.3) hold and remembering the relation between parameters $u$ and $v$ (1.6),
one can say that  our results   
 support   the conjecture  that the family of random graphs $\{\Gamma^{(n,\rho)}\}$
satisfies in the limit $(n,\rho)_\a\to\infty$ (1.9) a version of the weak graph theory Riemann Hypothesis 
\cite{HST,TS-III} that says that the Ihara zeta function 
$Z_\Gamma (u)$
is pole free for $\vert u\vert < 1/\sqrt q$, where $u\in \bC $ and $q+1$ is the maximum degree of all the vertices of $\Gamma$. In our setting, the maximum degree normalization is naturally replaced by
the averaged vertex degree $\rho$. Let us say that to prove this conjecture, one has to establish convergence
(3.2) for complex $v$ and this problem  goes also far beyond the frameworks of our results.

\vskip 0.3cm 
In the present note we  follow   a stochastic approach 
to the studies of the Ihara zeta function of graphs. 
We  assume the  graph's edges  to be present at random
and  consider the ensemble of such graphs  in the limit of their infinite dimensions. 
Such a reasoning   is completely in the spirit of the spectral theory 
of large random matrices 
used for the first time  in the spectral theory of heavy atomic nuclei 
\cite{P,W}, see also \cite{Me}. In the frameworks of the stochastic approach,
the collective  properties of complex systems remain valid while the difficulties are mostly related with
 special cases that are relatively rare.  This can be regarded as a kind of simplified description 
 that nevertheless catches the principal features of the system under consideration.

\vskip 0.1cm 
It should be noted that even in the frameworks of the    stochastic approach to IZF of graphs, 
the problem of the limiting transition 
of the infinitely increasing dimension of graphs, $n\to\infty$,
still persists.  The main difficulty in the establishing of convergence of the
normalized logarithm of IZF (1.4) is that  because of
possible presence of negative eigenvalues of $H^{(n,\rho)}$,
one can guarantee the existence of 
the  function $\Xi^{(n,\rho)}(v)$ (1.7) for small values of $\vert v\vert $ only,
and this smallness can converge to 0 as $n\to\infty$. However, the proportion of 
the graphs that exhibit such anomalously small area of convergence can be sufficiently weak. 
The fact  that the integral of the right-hand side of (3.5) converges confirms this conjecture.
We also see that due to (3.3), a kind of the averaged version of the the graph theory Riemann Hypothesis 
can be true for the  large irregular random graphs. 

\vskip 0.2cm

Let us finally point out  that statements  similar to our results could be obtained for the family 
of $d$-regular random graphs  $\hat \Gamma^{(n,d)}$. 
Regarding 
$
 \log Z_{\hat \Gamma^{(n,d)}}\left(u \right)$ (1.4)
with the spectral parameter $u= v/\sqrt{d-1}$, 
we get the following version of (3.1),
$$
\hat \Xi^{(n,d)} (v ) = {1\over n } \log \det \left( I(1+v^2) - {v\over \sqrt{d-1}} \hat A^{(n,d)}\right),
\eqno (3.6)
$$
where $\hat A^{(n,d)}$ is the adjacency matrix of $\hat \Gamma^{(n,d)}$.
With the help of the results of  \cite{McKay}, one can show 
that the right-hand side of (3.6)  converges as $n,d\to\infty$ for all $-1<v<0$ to the 
corresponding integral 
over  the shifted semicircle distribution (2.15). Here again, 
regarding  convergence of $\hat \Xi^{(n,d)} (v )$ in average, one can extend the above domain up to   $v\in (-1,1)$. 
To justify this, one would need to know the fine properties  of the graph's spectrum, 
including statements similar to the well-known  Alon's second eigenvalue conjecture (see e. g. \cite{F-1})
as well as estimates of the lowest eigenvalue of $\hat A^{(n,d)}/\sqrt{d-1}$ in the limit  $n,d\to\infty$.

\end{document}